\documentclass{amsart}

\usepackage{euscript}
  {}
  {}

\newtheorem{theorem}{Theorem}
\newtheorem{lemma}[theorem]{Lemma}

\newtheorem{maintheorem}{Main Theorem}

\theoremstyle{definition}

\theoremstyle{remark}

\newcommand{\abs}[1]{\left\vert#1\right\vert}

\newcommand{\R}{\mathbb R}

\newcommand{\nablash}{\nabla{\kern -.75 em
     \raise 1.5 true pt\hbox{{\bf/}}}\kern +.1 em}
\newcommand{\Deltash}{\Delta{\kern -.69 em
     \raise .2 true pt\hbox{{\bf/}}}\kern +.1 em}
\newcommand{\Rslash}{R{\kern -.60 em
     \raise 1.5 true pt\hbox{{\bf/}}}\kern +.1 em}

\newcommand{\gammab}{\bar\gamma}
\newcommand{\chib}{\bar\chi}

\newcommand{\Hb}{\bar H}

\newcommand{\Sphere}{\mathbb S}
\newcommand{\D}{\partial}

\title{Black hole initial data with a horizon of prescribed geometry}
\author{Brian Smith}

\begin{document}

\address{
  Freie Universit\"at Berlin, Arnimallee 6, 
  14195 Berlin, Germany}

\email{bsmith@math.fu-berlin.de}



\begin{abstract}
The purpose of this work is to construct asymptotically flat, time
symmetric initial data with an apparent horizon of   prescribed intrinsic 
geometry.  To do this, we use the parabolic partial differential
equation for prescribing scalar curvature. In this equation the
horizon geometry is contained within the freely specifiable part
of the metric.  This contrasts with the conformal method in which
the geometry of the horizon can only be specified up to a
conformal factor.
\end{abstract}

\maketitle

\section{Introduction}

In this work, the excision approach to constructing
black hole initial data is used. Specifically, black hole initial
data for the Einstein equations is taken to consist of the
following: (1) a manifold $M=\R^3\backslash K$, where $K$ is a
union of a finite number of balls. (2) a metric $g$ on $M$
asymptotic to the flat background metric $\delta$; in this work we
take this in the very strong sense that there  exist constants
$C_j$ for all $j\geq 0$ such that
$|r^{j+1}\D_j(g_{kl}-\delta_{kl})|\leq C_j$, where $r$ is the
coordinate radius.   (3) a second rank tensor field $k$ asymptotic
to 0. (4) a vectorfield $J$ asymptotic to 0. (5) a scalar field
$\rho$ asymptotic to 0. Furthermore, we require that
$(g,J,k,\rho)$ satisfy the  Einstein constraint equations
\begin{align*}
       R(g)+(\text{tr}k)^2-|k|^2_g&=16\pi\rho\\
     \nabla\cdot k-\nabla(\text{tr}k)^2&=-8\pi J,
\end{align*}
the dominant energy condition
\[
   \rho>|J|_g,
\]
and, in addition, an apparent horizon boundary condition on $\D
K$. To write the latter, let $n,H$ be the outward unit normal and
mean curvature of $\D K$, respectively.  The apparent horizon
boundary condition is that on $\D K$, one has
\begin{align*}
     \theta^+ &\equiv H+k(n,n)-\text{tr}k = 0\\
     \theta^- &\equiv -H+k(n,n)-\text{tr}k\leq 0.
\end{align*}
In addition, there should be no other compact surfaces 
satisfying $\theta^+\leq  0$.  

The interpretation of the data is that $M$ is a spacelike slice
with induced metric $g$ and second fundamental form $k$ in an
asymptotically flat spacetime.   The fields $J$ and $\rho$ are,
respectively, the local momentum current density and mass energy
density according to an observer comoving with $M$. The conditions
that we have assumed on $\D K$ imply, assuming weak or strong
cosmic censorship and the weak or strong energy condition, that if
the data were extended inside of $\D K$, this surface would form
the boundary of the \textit{totally trapped region} (for a
definition see~\cite{wald}), and thus $\D K$  is the
\textit{apparent horizon}.  The surface $\D K$ represents a black
hole within $M$ since, again, if $(M,g,k,J,\rho)$ were extended
inside $\D K$, this set would represent the outer boundary of the
largest set on $M$ which, based solely on the geometry of
 $(M,g,k,J,\rho)$, must be contained within the black hole region
of space time~\cite{HE},~\cite{wald}.  Thus, although $\D K$ is,
in general, not the set in which the spacelike slice intersects
the event horizon, within the initial data set itself it provides
the best possible representative of the black hole. In addition,
if the censorship and energy assumptions above hold, any data
inside $\D K$ will not affect the Cauchy development outside of
the event horizon, and for this reason we exclude it from
discussion. Finally, as shown by Hawking~\cite{H}, $\D K$  will
always consist of a finite union of topoligical spheres, and so
there has been no loss of generality in the topological
assumptions on $M$.

The problem of the construction of black hole initial data with
apparent horizon boundaries  has received much attention in recent
years. It would be impossible to cite every relevant work here,
but the interested reader may see, for
instance~\cite{D1},~\cite{D2},~\cite{D3},~\cite{C},\cite{M}. In
most of these works, the main tool for constructing black hole
initial data with an apparent horizon boundary has been the
conformal method, and considerable progress has been made.   This
method involves taking an initial``free'' metric on
$\R^3\backslash K$ and modifying it with a conformal factor to
ensure that the constraint equations are satisfied. The apparent
horizon condition appears as a Neumann boundary condition on $\D
K$.  However, with this method one has no control over the values
of the conformal factor on the horizon; it is therefore impossible
to specify the horizon geometry with this method.

Since in the present work we want to prescribe the  geometry of
the apparent horizon, we use instead the parabolic equation for
prescribing scalar curvature, for which the horizon geometry will
be contained in the freely specifiable part of the metric.  In~\cite{bartnik93}, Bartnik does this for quasi-spherical metrics, 
which allows for the construction of
apparent horizons that are intrinsically round. In the present
work we relax this condition. Since this is a first attempt in
that direction we are also going to make the simplifying
assumptions that the horizon only has one component,  the initial
data set has moment of time symmetry $k=0$, and the local mass
density $\rho$ is compactly supported. This reduces the problem to
the following: \textit{Given a compactly supported local mass
density $\rho\geq 0$ on $\R^3\backslash B_{r_0}(0)$, and a metric
$h$ on $\D B_{r_0}(0)\equiv \Sphere^2$, construct a metric $g$ on
$\R^3\backslash B_{r_0}(0)$, asymptotic to the standard flat
metric $\delta_{ij}$, satisfying $R(g)=16\pi\rho$, and such that
$g|_{\D B_{r_0}(0)}=h$.}

To attack this problem  we make the
standard spherical-polar identification of $M$ with
$[r_0,\infty)\times\Sphere^2$ and construct $g$ in the form
\begin{equation}\label{eq:formofmetric}
     g=u^2dr^2+r^2\gamma,
\end{equation}
where $\gamma$ vanishes on $\D_r$ and $g|_{S_r}=r^2\gamma$. The
parabolic scalar curvature equation, which we write in the next
section, may be viewed as a second order parabolic partial
differential equation for the radial component $u$ of the metric as long as 
$\gamma^{AB}r\D_r\gamma_{AB}>-4$. 
The component $\gamma$ appears in the coefficients along with the
scalar curvature $R$.

To see how we can obtain the apparent horizon condition for a
metric written in the form~\eqref{eq:formofmetric} note that the
Schwarzschild  metric cast into this form reads
\[
         g_s=\frac{1}{1-\frac{r_0}{r}}dr^2+r^2\gamma,
\]
where $\gamma$ is a fixed round metric on $\Sphere^2$.  Of course,
at first glance it appears that the metric is singular at $r_0$,
but it is well known that an appropriate coordinate transformation
shows that $g_s$ can actually be extended to the boundary as a
smooth metric on a manifold with boundary.  The blow-up in the
normal component is, in fact, responsible for the surface
$r=r_0$ being an apparent horizon for this metric. Indeed, this
follows since the extrinsic curvature of the foliation spheres of
a metric in the form~\eqref{eq:formofmetric} is
\[
       \chi=\frac{r}{u}\left(\gamma+\frac{1}{2}r\frac{\D\gamma}{\D
       r}\right),
\]
and so for the Schwarzschild metric one has $\chi=\sqrt{r}\sqrt{r-r_0}\gamma.$
Of course, in this case we actually know that $S_{r_0}$ is a
spacelike slice of the event horizon.

Our approach to obtaining a horizon at $r=r_0$ will be to imitate
the Schwarzschild case as closely as possible  while striving for
the greatest generality in the constructed metric.   Namely, we
shall construct metrics in the form
\[
    g_s=\frac{v^2}{1-\frac{r_0}{r}}dr^2+r^2\gamma
\]
by choosing $\gamma$ at the outset and  using the parabolic scalar
curvature equation to solve for $v$  in order to get
$R(g)=16\pi\rho\geq 0$.  Then the horizon geometry is just given
by $r^2\gamma$ and is thus contained within the freely specifiable
part of the metric. We may implement this procedure for a large
class of $\gamma$.  In particular, one has
\begin{maintheorem}\label{thm:initialdata}
Let $h$ be a metric on $\Sphere^2$ with positive Gauss curvature.
Let $\varepsilon>0$ and let  $\gammab$ denote the fixed round metric on $\Sphere^2$.
Let $\gamma(r)$ be  a smooth family of metrics
on $[r_0,\infty)$ that satisfies, for some large $r_2>r_0+\varepsilon$,
the following:
\begin{align*}
       r_0^2\gamma(r_0)&=h\\
       \gamma(r)&\equiv \gamma(r_0),\,r\in [r_0,r_0+\varepsilon)\\
       \gamma(r)&\equiv\gammab,\, r>r_2\\
       r\frac{\D\gamma_{AB}}{\D r}\gamma^{AB}&>-4\\
       \kappa(\gamma)&>0
\end{align*}
Then for a compactly supported function  $\rho$ on
$M=[r_0,\infty)\times\Sphere^2$ satisfying $16\pi\rho<\kappa$, and
which  also vanishes on $[r_0,r_0+\varepsilon)$, there exists
asymptotically flat time symmetric initial data $(g,0)$ on $M$ of
mass density $\rho$ such that $g$ has the form
\[
     g=\frac{v^2}{1-\frac{r_0}{r}}dr^2+r^2\gamma,
\]
where $v$ is $C^{\infty}$, independent of $r$ on
$[r_0,r_0+\varepsilon)$, and bounded above and below by positive
constants. The metric $g$ has an apparent horizon at $r=r_0$ with
intrinsic geometry $(\Sphere^2,h)$; the horizon is, in fact,
totally geodesic.
\end{maintheorem}
 This theorem is an immediate corollary of a slight generalization
to be proved at the end of the last section, in which we replace
the conditions $\kappa>0,16\pi\rho<\kappa$ with a  more general,
but  more complicated, integral condition involving
$\rho,\Hb,A,\kappa,\kappa|_{r_0}$.

A few remarks are also in order  concerning the other conditions
on the free part of the metric $\gamma$. The first of these is
obviously what allows us to specify the horizon geometry.  The
second is a technical condition, but only affects the generality
of $\gamma$ on the region $[r_0,r_0+\varepsilon]$, which can be
made arbitrarily small;  we shall refer to this region
$[r_0,r_0+\varepsilon]\times\Sphere^2$ as the \textit{collar
region}. The third is a technical restriction that can be removed
and replaced with an assumption on the decay of $\gamma$ to
$\gammab$.  However, doing so would require a detailed analysis of
the asymptotic behavior, and this would detract from the purpose
of the current work.    The next to  last condition serves the 
technical purpose that it causes the parabolic scalar curvature 
equation to be uniformly parabolic for $r$ strictly greater than $r_0$, but it also   
implies that  the mean curvature 
of the foliation will be positive
since the mean curvature of these spheres can, in general, be
calculated to be
\[
      H=\frac{1}{ru}\left(2+\frac{1}{2}r\frac{\D\gamma_{AB}}{\D
      r}\gamma^{AB}\right).
\]
This fact  ensures that outside of $r=r_0$
there are no other compact surfaces satisfying $\theta^+\leq 0$,
and so $r=r_0$ is an apparent horizon.  

The outline of the paper is as follows: In the next section we write 
the parabolic scalar curvature equation and present some necessary results 
concerning this equation that were obtained in previous works.  In the last 
section we prove the slight extension of our main theorem in two steps.  
In the first step we construct the metric on the collar region by solving 
an elliptic equation for $v$ that we obtain from the parabolic scalar 
curvature equation by separation of variables;  the conditions on the collar 
region are precisely what allows us to do this.  This step is used to 
overcome the fact that the parabolicity of the parabolic scalar 
curvature equation must break down on the horizon.    
In the second step, which is contained in the proof of Theorem~\ref{thm:lastthm}, 
the metric exterior to the collar
region is constructed using the results from the first section.

\section{The parabolic scalar curvature equation}
The proof of the main theorem~\ref{thm:initialdata} is essentially a
problem in prescribed scalar curvature.  That is, given the
``free" part of the metric $\gamma$, and the function $\rho$, we
want to choose the remaining component $u$ such that the metric
$g=u^2dr^2+r^2\gamma$ has scalar curvature $R=16\pi\rho$.  To do
this, we use the parabolic scalar curvature equation, which
relates $u$ to $R,\gamma$:
\begin{equation}\label{eq:primus}
    \Hb r\frac{\D u}{\D r}=u^2\Delta_{\gamma}u+Au-\left(\kappa-\frac{r^2R}{2}\right)u^3,
\end{equation}
where $\kappa|_r$ is the Gauss curvature of $\gamma(r)$ and
\begin{gather*}
   A=r\frac{\D\Hb}{\D r} - \Hb +\frac12 \abs{\chib}_\gamma^2 + \frac12 \Hb^2, \\
   \chib_{AB}=\gamma_{AB}+\frac 12 r\frac{\D\gamma_{AB}}{\D r}, \\
   \Hb=\text{tr}_{\gamma}\chib=2+\frac 12 r\frac{\D\gamma_{AB}}{\D r}\gamma^{AB};
\end{gather*}
$A,B$ are used to denote components with respect to local
coordinates $(\theta^1,\theta^2)$ on $\Sphere^2$. For derivations of this equation 
see \cite{bartnik93},~\cite{SW},~\cite{ST}.  The equation has also been
used successfully in~\cite{ST2}~\cite{WY}.  For a discussion of the 
use of this equation to construct non-time-symmetric initial data 
see~\cite{bartnik91},~\cite{bartnik03},~\cite{Sh}.

Note that if
$\Hb>0$ and $u$ is bounded above and below by positive constants
then this equation is uniformly parabolic. As already noted, this
is equivalent to the positivity of the mean curvature of the
foliation spheres. Of course, this requires some pointwise a
priori bounds on the solution $u$, but in the case that $u$ is
positive and bounded initially,  these are easily obtained by the
maximum principle under appropriate assumptions on the free data.
The result of this is contained in the next lemma, whose proof can
be found in~\cite{SW},~\cite{SW2}, and is a slight generalization
of a result contained in~\cite{bartnik93}.  The proof is 
repeated here since it yields simple but important bounds on the
components of the constructed metric. To state this, given a
function $f$ on $[r_1,r_2]\times\Sphere^2$ put
$f^*(r)=\sup_{p\in\Sphere^2}f(r,p)$,
$f_*(r)=\inf_{p\in\Sphere^2}f(r,p)$.
\begin{lemma}
For $r\in I\equiv [r_1,r_2]$, let $\gamma(r)$ be a family  of
metrics on $\Sphere^2$ and let $\Hb,\kappa,A,R$ be defined in
terms of $\gamma$ as above. Define
\[
       K_I=\sup_{r\in I}\left\{\frac{1}{r_1}\int_{r_1}^r
       \left(\frac{2}{\Hb}\left(\frac{r^2R}{2}-\kappa\right)\right)^*(s)
       \,\exp\left(\int_{r_1}^s\left(\frac{2A}{\Hb}-1\right)^*\frac{dt}{t}\right)ds \right\}.
\]
Assume $\Hb>0$ and $K<\infty$.  Suppose that $u$ is a solution of equation~\eqref{eq:primus}  
on $[r_1,r_2]$ such that   ``initially" one has $0<u(r_1,\cdot)<1/\sqrt{K}$.
Then  there exists a
constant $C$ depending on $\gamma, R, u_1, I$ such that 
$ C^{-1}<u< C$.
\end{lemma}
\begin{proof}
In order to prove the result, it is useful to introduce  the auxiliary
function $w=u^{-2}$ to convert Equation~\eqref{eq:primus} into an
equation which is linear everywhere but the principal part:
\[
  r\D_rw=-\frac{2}{\Hb}\frac{\Delta u}{u}-\frac{2A}{\Hb}w-\frac{2}{\Hb}\left(\frac{r^2R}{2}-\kappa\right)
\]
By using the maximum principle (see~\cite{SW2})  one  can bound
a solution of this equation with initial data $w_1$ from below by a
solution of the ordinary differential equation
\[
      r\frac{dw_*}{dr}=-\left(\frac{2A}{\Hb}\right)^*w_*-\left(\frac{2}{\Hb}\left(\frac{r^2R}{2}-\kappa\right)\right)^*
\]
with any initial data satisfying $w_*(r_1)\leq w(r_1,\cdot)$.  For convenience
put $a=2A/\Hb, b=r^2R/2-\kappa$ so that the previous equation can
be written $rw'+a^*w=-b^*$.  Using the integrating factor
$\exp\int_{r_1}^r(a^*-1)/t\,dt$ converts the equation into the form
\[
    \frac{d}{dr}\left(r\exp\int_{r_1}^{r}\left(\frac{a^*-1}{t}\right)dt\,w_*\right)
    =-b^*\exp\int_{r_1}^{r}\left(\frac{a^*-1}{t}\right)dt,
\]
which can be immediately integrated to yield
\[
      w_*=\frac{1}{r}\exp\int_{r_1}^{r}\left(\frac{1-a^*}{t}\right)dt
      \left\{r_1w_*(r_1)
      -\int_{r_1}^rb^*\exp\int_{r_1}^{s}\left(\frac{a^*-1}{t}\right)dt\,ds\right\}.
\]
The hypothesis of the lemma ensures that on the interval $I$ the right hand side will
be bounded below by a positive constant. This in turn gives the
upper bound for $u$.  To obtain the lower bound for $u$, we note
that by using the maximum principle again $w$ can be bounded
above by solutions of
\[
     r\frac{dw^*}{dr}=-\left(\frac{2A}{\Hb}\right)_*w^*-\left(\frac{2}{\Hb}\left(\frac{r^2R}{2}-\kappa\right)\right)_*
\]
satisfying $w^*(r_0)>w(r_0,\cdot)$.
\end{proof}
If, in addition, the coefficients of Equation~\eqref{eq:primus}
are $C^{\infty}$ then the bounds on $u$ just obtained guarantee
existence of a $C^{\infty}$ solution with the initial data $u_1$.
This is carried out in~\cite{bartnik93},~\cite{SW}.  The result is
contained in the next theorem.
\begin{theorem}  \label{thm:exist}
Suppose that the hypotheses of the previous lemma are satisfied
for $\gamma,R\in C^{\infty}([r_1,r_2]\times\Sphere^2)$ and $u_1\in
C^{\infty}(\Sphere^2)$. Then there exists a unique $C^{\infty}$ solution $u$ of
Equation~\eqref{eq:primus}  on $I\equiv [r_1,r_2]$ with the initial
data $u(r_1,\cdot)=u_1$.  Furthermore $ C^{-1}<u< C$,
where $C$ is the constant from the previous lemma.
\end{theorem}
Note that in the previous theorem $r_2$ can be as large as we
would like so that, in fact, we get existence as $r\to\infty$ as
long as the free data continue to satisfy the hypotheses of the theorem on any interval.
 In our case we would actually
like to obtain an asymptotically flat metric as a consequence of
this.  Hence, we need to assume appropriate asymptotic behavior on
$\gamma,R$, which in our case we have done, and derive from this
the correct asymptotic behavior of the solution $u$.  As in
previous work for Equation~\eqref{eq:primus},
 this is done by proving the
boundedness of the function $m$ defined by
\[
      u^{-2}=1-\frac{2m}{r}.
\]
The function $m$ verifies the equation
\[
    \Hb r\D_{r}m=
      \frac{\Delta_{\gamma}u}{u}
     -\left(2A-\Hb\right)m+{r}(A-B).
\]
Since we have made the  assumptions that on a
neighborhood of infinity $\gamma$ is identically the round metric
and $R$ vanishes, the equation simplifies
dramatically.  Indeed, we have $\Hb\equiv 2,A=1$ so that the
equation for $m$ becomes
\begin{equation}\label{eq:meq}
       2 r\D_{r}m=
      \frac{\Delta_{\gamma}u}{u}.
\end{equation}
Bartnik has made a detailed analysis of this equation in~\cite{bartnik93},
in which the following is proved:
\begin{theorem}
Suppose that $m$ is defined and $C^{\infty}$ on $[r_0,\infty)\times\Sphere^2$
and satisfies Equation~\eqref{eq:meq}, where $m,u$ are related as  above.
Then there is a constant $m_0$ and $\varphi\in\Sphere^2$ verifying $(\Delta+2)\varphi=0$ such that
\[
     m=m_0+\frac{\varphi}{r-2m_0}+\epsilon(r),
\]
where $\epsilon$ is a $C^{\infty}$ function on
 $[r_0,\infty)\times\Sphere^2$ satisfying,
\[
     |(r\D r)^i\nabla^j\epsilon|\leq C_{i,j}r^{-3}\log r,
\]
for all $i,j\geq 0$ such that $i+j\leq k$.
\end{theorem}
This theorem shows, in particular, that the constructed metric $g$
will be asymptotically flat in the sense mentioned in the introduction.
Furthermore, the mass of the manifold is $m_0$.

Putting the previous results together, we get the result needed to
construct the metric outside of the collar region.
\begin{theorem} \label{thm:awayfromcollar}
For $r\in [r_1,\infty)$ let $\gamma(r)$ be a family of
metrics on $\Sphere^2$ satisfying $\Hb>0$ and $K<\infty$,
where $K=K_{[r_0,\infty)}$ is defined as in Theorem~\ref{thm:exist}.  Furthermore,
assume  for some $r_2>r_1$ that $\gamma$ is the fixed round
metric on $\Sphere^2$ and $R\equiv 0$ for $r>r_2$.
Then if $u_1\in C^{\infty}(\Sphere^2)$ satisfies $0<u_1<1/\sqrt{K}$,
there exists a unique $C^{\infty}$ asymptotically flat metric $g$ in the form
\[
           g=u^2dr^2+r^2\gamma,
\]
with $u(r_1,\cdot)=u_1$, and whose scalar curvature satisfies $R(g)=R$.
\end{theorem}

\section{constructing the data}

To construct the black hole initial data we  construct the data on the collar region,
and make sure this patches smoothly with data constructed  via Theorem~\ref{thm:awayfromcollar}
for $r>r_1$.  At first glance, it seems problematic to use
Equation~\eqref{eq:primus} to solve for $u$ on the collar region since at
$r=r_0$  the equation must fail to be parabolic if this surface is to be a horizon.  This
problem is resolved by assuming that $\gamma$ is fixed on this region.
 Assuming in
addition that $v$ is also fixed in $r$ on the collar region yields
an elliptic equation for $v$ on $\Sphere^2$. Indeed, the fact that
$\gamma$ does not vary in $r$ immediately gives that $\Hb\equiv
2,A\equiv 1$ so that Equation~\eqref{eq:primus} becomes
\[
      2 r\frac{\D u}{\D
      r}=u^2\Delta_{\gamma}u+u-\kappa u^3.
\]
Substituting $u=v\left(1-r_0/r\right)^{-\frac{1}{2}}$ yields
the following equation for $v$:
\begin{equation}\label{eq:vellipt}
     \Delta_{\gamma}
     v-\kappa v+\frac{1}{v}=0.
\end{equation}
One has
\begin{theorem}  \label{thm:vellipt}
Assume $\kappa>0$ and let $\kappa_*=\inf_{\Sphere^2}\kappa, 
\kappa^*=\sup_{\Sphere^2}\kappa$. Then 
Equation~\eqref{eq:vellipt} has a  
positive solution $v\in C^{\infty}$ 
satisfying $1/\sqrt{\kappa^*}\leq v\leq 1/\sqrt{\kappa_*}$. 
Within the class of $C^2$ functions there are exactly two solutions $\pm v$.
\end{theorem}
\begin{proof}
We are going to produce the solution by the method of
sub-solutions. Before starting the process, we note that
sub(super)-solutions for Equation~\eqref{eq:vellipt} in the sense of
\[
       \Delta_{\gamma}
     v-\kappa v+\frac{1}{v}\geq(\leq) 0
\]
are also sub-solutions  and super-solutions in the sense that any
super-solution bounds any sub-solution from above. As usual, this is
just a consequence of the maximum principle, but we should
carefully check this anyway: Assuming that $v_*, v^*$ are sub and
super solutions of Equation~\eqref{eq:vellipt} in the sense of the
previous inequality, we consider the equation for the difference
$\delta v^*=v^*-v_*$
\[
    \Delta\delta v^*-(\kappa+\frac{1}{v_*v^*})\delta v^*\leq 0.
\]
To see that $\delta
v^*\geq 0$, suppose instead that $\delta v^*<0$ somewhere, and
consider the set $\Omega\equiv \{p\in\Sphere^2:\delta v^*(p)<0\}$.
Since the term in parentheses is positive, and since we have
assumed $\Omega$ is nonempty, by the maximum principle one has
$\inf_{\Omega}\delta v^*=\inf_{\D\Omega}\delta v^*$, which is a
contradiction since $\delta v^*=0$ on $\D\Omega$.

To now begin the method of sub-solutions, we are going to take as
our starting sub-solution $v_*=1/\sqrt{\kappa^*}$.  For future reference,
note that $v^*=1\sqrt{\kappa_*}$ is a super-solution.
Choose
$\lambda>1/v_*^2$, and define $v_1$ as the solution of
\[
     \Delta v_1-(\lambda+\kappa)v_1=-\lambda v_* -\frac{1}{v_*}.
\]
Then an application of the maximum principle just as in the first paragraph
shows that $v_1>v_*$ since the difference $\delta v_1=v_1-v_*$ satisfies
\[
         \Delta\delta v_1-(\lambda+\kappa)\delta v_1\leq 0.
\]

We now define $v_i$ inductively as the solutions of
\begin{equation}\label{eq:recursive}
          \Delta v_i-(\lambda+\kappa)v_i=-\lambda v_{i-1} -\frac{1}{v_{i-1}}.
\end{equation}
The fact that $v_i\geq v_{i-1}$ now follows from induction and
another application of the maximum principle  since  $v_{i-1}\geq
v_{i-2}$ yields for the difference $\delta v_i=v_i-v_{i-1}$ the
inequality
\[
      \Delta\delta v_i-(\lambda+\kappa)\delta v_i=-\left(\lambda-\frac{1}{v_{i-1}v_{i-2}}\right)\delta v_{i-1}\leq 0.
\]

Thus, we now have an increasing sequence of functions $v_i$, which
are, in fact sub-solutions since for the right hand side
of~\eqref{eq:recursive} we have
\[
         -\lambda v_{i-1} -\frac{1}{v_{i-1}}
    = -\lambda v_{i} -\frac{1}{v_{i}}+\left(\lambda-\frac{1}{v_{i}v_{i-1}}\right)(v_i-v_{i-1})
       \geq -\lambda v_{i} -\frac{1}{v_{i}}.
\]
Hence, we now have a pointwise non-decreasing sequence of
sub-solutions $v_i$ that satisfies $v_*\leq v_i\leq v^*$. 
Thus $v_i$ converges to some function $v$ pointwise.  
In addition, the bounds on $v_i$ imply that the right hand side of 
the Equations~\eqref{eq:recursive} are uniformly bounded in $L^q$ for any $q$.
Whence, by applying elliptic regularity theory 
we see that the $v_i$ are also uniformly bounded in $W^{2,q}$.  
Applying elliptic regularity iteratively, we may bootstrap to see that, 
in fact, the $v_i$ are uniformly bounded in $W^{k,q}$ for any $k,q$ so
that the Sobolev embedding theorem shows that they are uniformly
in $C^k$ for any $k$. The Ascoli-Arzela theorem then shows that a
subsequence converges in $C^k$ to a function that can be none
other than $v$.  Taking the limit in Equation~\eqref{eq:recursive}
then shows that $v$ must be a solution of
Equation~\eqref{eq:vellipt}.

Concerning the uniqueness, first note that if a solution is to be of class $C^2$, 
then it cannot change signs, and so we may restrict to the class of positive functions.  
Given two positive solutions $v_1,v_2$ one can see that $v_1=v_2$ by, 
again, applying the maximum principle to the equation for the difference $\delta v=v_2-v_1$
\[
   \Delta\delta v-\left(\kappa+\frac{1}{v_1v_2}\right)\delta v=0.  
\]    
\end{proof}
We may now prove our  final theorem, of which the main theorem  
is an immediate corollary.
\begin{theorem}     \label{thm:lastthm}
Let $\gamma$ be a family  of metrics as in
the hypothesis of the main theorem with the exception that we do not
assume $\kappa(\gamma)>0$ outside of the collar region. Instead,
assume that $\gamma, \rho$ are such that, with $R=16\pi\rho$ and
$K$ defined as in Theorem~\ref{thm:awayfromcollar}, one has
\[
        K<r_0^2\kappa(h)\left(1-\frac{r_0}{r_1}\right).
\]
Then there exists
asymptotically flat time symmetric initial data $(g,0)$ on $M$ of
mass density $\rho$ such that $g$ has the form
\[
     g=\frac{v^2}{1-\frac{r_0}{r}}dr^2+r^2\gamma,
\]
where $v$ is $C^{\infty}$ and bounded above and below by positive
constants.
\end{theorem}
\begin{proof}
From the previous theorem  we have the metric in the above form on
the collar region. Taking $u_1=v/\sqrt{1-r_0/r_1}$ in
Theorem~\ref{thm:awayfromcollar} gives the metric for $r>r_1$
since $0<u_1<1/\sqrt{K}$. We only have to check that the
constructed metric is smooth across $r=r_1$.  This is easily done
by applying Theorem~\ref{thm:awayfromcollar} again for $r\in
[r_1-\delta,\infty)$ for very small $\delta$.  Indeed, the new
metric is certainly $C^{\infty}$ at $r=r_1$ and agrees with the
previous metric on $[r_1-\delta,\infty)$ by the uniqueness part of
that theorem.
\end{proof}

\bibliographystyle{amsplain}

\begin{thebibliography}{99}

\newcommand{\aut}[1]{{\sc #1},}
\newcommand{\tit}[1]{{\em #1\/},}
\newcommand{\vol}[1]{{\bf #1}}
\newcommand{\yr}[1]{(#1)}
\newcommand{\pp}[2]{#1--#2}



\bibitem{bartnik93}
    \aut{R.~Bartnik}
    \tit{Quasi-spherical
         metrics and prescribed scalar curvature}
    J. Differential Geom.
    \vol{37} No.~1
    \yr{1993}
    \pp{31}{71}.

\bibitem{bartnik91}
    \aut{R.~Bartnik}
    \tit{Initial data for the Einstein equations in the quasi-spherical gauge}
    Gravitation and Astronomy Instrument Design and Astrophysical Prospects, edited by D. McLelland and H.A. Bachor, World Scientific, Singapore, 1991
    
\bibitem{bartnik03}
    \aut{R.~Bartnik, J. Isenberg}
    \tit{The constraint equations} gr-qc/0405092
    \yr{2003} gr-qc/0405092
    

\bibitem{D1}
    \aut{S. Dain}
   \tit{Trapped surfaces as boundaries for the constraint equations}
    Class. Quantum Grav.
    \vol{21}
    \yr{2004}
    \pp{555}{573}.

\bibitem{D2}
    \aut{S. Dain, J.L. Jaramillo, and Badri Krishnan}
   \tit{On the existence of initial data containing islolated black holes}

    \vol{}
    \yr{2004} gr-qc/0412061


\bibitem{D3}
    \aut{S. Dain}
   \tit{On black holes as inner boundaries for the constraint equations}

    \vol{}
    \yr{2006} gr-qc/0401018


\bibitem{C}
    \aut{Gregory B. Cook}
   \tit{Initial data for Numerical Relativity}
    Living Reviews in Relativity, www.livingreviews.org
    \vol{3}
    \yr{2000}





\bibitem{H}
    \aut{S.W. Hawking}
    \textit{The Event Horizon} In \textit{Black Holes}
    Les Houches Lectures (1973), edited by C. DeWitt, B.S. DeWitt
     Amsterdam: North Holland,1972





\bibitem{HE}
    \aut{S.W. Hawking and G.F.R. Ellis}
    \textit{The Large Scale Structure of Spacetime},
    Cambridge University Press, Cambridge, 1973.






\bibitem{lieberman}
    \aut{G. Lieberman}
    \textit{Second Order Parabolic Partial Differential Equations},
    World Scientiic, New Jersey, 1996.


\bibitem{M}
    \aut{D. Maxwell}
   \tit{Solutions of the Einstein constraint equations with apparent horizon boundaries}
    \yr{2004} gr-qc/0307117


\bibitem{Sh}
    \aut{J. Sharples}
   \tit{Local existence of quasispherical space-time initial data}
    Journal of mathematical Physics 
    \vol{46} 
    \yr{2005}

\bibitem{ST}
    \aut{Y.~Shi and L.~Tam}
   \tit{Positive mass thoerem and the boundary behaviors of compact
    manifolds with nonnegative scalar curvature}
    Journal of Differential Geometry
    \vol{62} No. 1
    \yr{2002}
    \pp{79}{125}.


\bibitem{ST2}
    \aut{Y.~Shi and L.~Tam}
   \tit{Quas-local mass and the existence of horizons}


    \yr{2005} DG/0511398




\bibitem{SW}
    \aut{B.~Smith and G.~Weinstein}
    \tit{On the connectedness of the space of initial data for the
         Einstein equations}
    Electron.\ Res.\ Announc.\ Amer.\ Math.\ Soc.
    \vol{6}
    \yr{2000}
    \pp{52}{63}.




\bibitem{SW2}
    \aut{B.~Smith and G.~Weinstein}
    \tit{Quasi-convex foliations and asymptotically flat metrics of non-negative scalar curvature}
    Communications in Analysis and Geometry
    \vol{12} No.~3
    \yr{2004}
    \pp{511}{551}.



\bibitem{wald}
    \aut{Robert M. Wald}
    \textit{General Relativity}
    The University of Chicago Press, Chicago, 1984.


\bibitem{WY}
    \aut{Mu-Tao Wang and Shing-Tung Yau}
   \tit{A generalization of Liu-Yau's quasi-local mass}

    \vol{}
    \yr{2006} DG/0602321


\end{thebibliography}

\end{document}